\documentclass[12pt, preprint]{aastex}

\usepackage{graphicx} \usepackage{natbib}

\slugcomment{To appear in The Astrophysical Journal}

\shorttitle{Debris Around HD69830} \shortauthors{Beichman etal.}

\begin{document}

\title{An Excess Due to Small Grains Around The Nearby K0V Star HD69830: Asteroid or Cometary Debris?}

\author{C. A. Beichman}

\affil{Michelson Science Center, California Institute of Technology, M/S 100-22, Pasadena, CA, 91125}

\email{chas@pop.jpl.nasa.gov}

\author{G. Bryden, T. N. Gautier, K. R. Stapelfeldt, M. W. Werner}

\affil{Jet Propulsion Laboratory, California Institute of Technology}

\author{ K. Misselt, G. Rieke, J. Stansberry, D. Trilling}

\affil{Steward Observatory, Univ. of Arizona}

\begin{abstract}

Spitzer photometry and spectroscopy of the star HD69830 reveal an excess of emission relative to the stellar photosphere between 8 and 35~$\mu$m dominated by strong features attributable to crystalline silicates with an emitting surface area more than 1,000 times that of our zodiacal cloud. The spectrum closely resembles that of the comet C/1995 O1 (Hale-Bopp). Since no excess is detected at 70~$\mu$m, the emitting material must be quite warm, be confined within a few AU of the star, and originate in grains with low long-wavelength emissivity, i.e. grains much smaller than $70\, \mu$m$/2\pi\sim 10 $~$\mu$m. The strong mineralogical features are evidence for even smaller, possibly sub-micron sized grains. This small grain size is in direct contrast to the 10-100~$\mu$m grains that dominate the relatively featureless spectra of our zodiacal dust cloud and most other main sequence stars with excesses. The upper limit at 70~$\mu$m also implies that any Kuiper Belt analog must either be very cold or be less massive than $\sim$ 5 times our own Kuiper Belt.

With collisional and Poynting-Robertson drag times of about a thousand years for  small grains, the emitting material must either: a) be created through continual grinding down of material in a dense asteroid belt; or b) originate in cometary debris arising from  either a single ``super-comet'' or a very large number of individual comets arriving from a distant reservoir. In the case of a cometary origin for the emission, the mass requirements for continuous generation by many individual comets are unreasonable and we favor the capture of a single ``super comet'' into a 0.5-1 AU orbit where it can evolve a large number of small grains over a  2 Myr period.

\end{abstract}


\section{Introduction}

The debris disk phenomenon, first discovered by the Infrared Astronomy Satellite (IRAS) (Aumann et al. 1984; Gillett 1986; Backman \& Paresce 1993) is of great interest because of the clues these disks offer toward understanding the formation and evolution of planetary systems. Toward this end, aspects of debris disks have been studied with the Infrared Space Observatory (ISO) (Habing et al. 2001; Decin et al. 2003), optical imaging (e.g., Smith \& Terrile 1984; Heap et al. 2000), and in the submillimeter (Holland et al. 1998; Greaves et al. 1998; Dent et al. 2000).

Information on the nature of the dust grains in these disks is particularly important. Spectroscopic observations toward luminous Herbig Ae/Be stars such as HD100546 \citep{grady1997, malfait1998, meeus2001} and in young debris stars such as $\beta$ Pictoris (Telesco and Knacke 1991) and 51 Oph (Fajardo-Acosta et al. 1993) revealed dust composed, at least in part, of small (sub-micron) grains of crystalline silicates such as forsterite and enstatite. The similarity of these spectral features to those seen in comet C/1995 O1 (Hale-Bopp) 
\citep{crovisier1996, grun2001, wooden2000} suggests that this circumstellar material may represent cometary debris. Yet these spectral features are not present in all or even most debris disks. More than a dozen classic debris disks (including Fomalhaut) examined by Spitzer \citep{Jura2004, stapelfeldt2004} show little or no spectral structure, suggesting that the grains in these systems are larger than $\sim$10 $\mu$m. These grains may be similar to those in our own zodiacal cloud which are predominantly larger than 10-100~$\mu$m with only a small admixture of smaller silicate grains yielding a weak 10~$\mu$m emission feature \citep{kelsall1998, fixsen2002, reach2003}. 

In this paper we report the detection of a large excess due to hot grains orbiting the star HD69830. These grains are dominated by a population of crystalline silicates with prominent features in the IRS spectrum.

\section{Properties of HD69830}

HD69830 (Gliese 302, BD-12 2449, HR 3259) is a bright (V=5.95 mag), nearby (12.6 pc), main sequence dwarf of solar metallicity, [Fe/H]=0.015 \citep{destrobel2001}. Its spectral type has been variously classified as between G8 \citep{skiff2003} and K0 (Song et al. (2000) and other references in SIMBAD). We adopt a K0V spectral type throughout this paper which implies an effective temperature of 5150 K, a mass of 0.8 M$_\odot$, and a total luminosity of 0.45 L$_\odot$ \citep{allen2000}. Using isochrone fitting, Li absorption, Ca II H\&K line activity, weak X-ray emission, and space motions, Song et al. (2000) suggested an age older than 0.6 Gyr and up to 2 Gyr with a preference for the older age. Examination of Ca II H\&K activity indices leads Marcy and Fischer (private communication, 2005) and Wright et al. (2004) to prefer an age of 2-4.7 Gyr while IUE spectra suggest an age of 3 Gyr \citep{hufnagel1994}. We adopt 2 Gyr as the probable age for this star. 

No planets have been detected orbiting HD69830 using the radial velocity technique with a limit on any planet with a mass, M sin(i) $>$ 0.5 M$_{Jup}$ located within a distance of 3 AU of the star (Marcy and Fischer, private communication). Although Mannings and Barlow (1998) noted the presence of a small ``hot" excess based on IRAS observations at 25~$\mu$m (3.8$\sigma$, Table~\ref{phottable}), IRAS did not detect the star at either 60 or 100~$\mu$m. The star was not observed with ISO.

\section{Spitzer Observations}

\subsection{Observations and Data Reduction}

Spectra of HD69830 were obtained with {\it Spitzer} on 2004 April 18 (AOR 4016640). These measurements were part of a program that has observed 36 main-sequence stars to look for disks (Beichman et al. 2005; Bryden et al. 2005) using the Infrared Spectrograph (IRS) (Houck et al. 2004). Standard IRS Staring mode observations were made with the Short Low Order 1 (SL1; 7-14~$\mu$m), Long Low Order 2 (LL2; 14-20~$\mu$m), and Long Low Order 1 (LL1; 20-35~$\mu$m) modules. Each star was observed at two positions along the slit, called Nod$_1$ and Nod$_2$. Data were processed by the Spitzer Science Center (SSC) to produce calibrated images of the spectrometer focal plane. The SSC SPICE software was used to extract spectra from these images. Because of the need for careful subtraction of the stellar continuum to detect a faint excess, additional steps were taken in producing the final spectra:

\begin{itemize}

\item To ensure proper sky subtraction, differences were formed between the two Nod positions to produce two sky-subtracted images (Diff$_1$=Nod$_1-$Nod$_2$ and Diff$_2$=Nod$_2-$Nod$_1$). SPICE was used to extract 1-$D$ spectra from each difference image using default settings and calibration.

\item The first few points at the beginning and end of the spectrum from each module were typically unreliable, as were a few bad pixels flagged in the SSC processing. These were rejected. This effect was particularly noticeable at the long wavelength end of SL1 ($\lambda>14\, \mu$m). Since the short wavelength end of LL2 overlaps SL1, there is no gap in the final spectrum.

\item Longward of 14~$\mu$m, photospheres of solar type stars are smooth and do not differ greatly from a Rayleigh-Jeans blackbody \citep{kurucz1992, castelli2003}. We used this fact to improve the pixel-to-pixel calibration in the extracted SSC/SPICE spectra. 35 of the 36 stars observed with IRS as part of this program showed no evidence for an excess at either 24 or 70~$\mu$m with MIPS, but did show consistent pixel-to-pixel deviations from a smooth photospheric model. From these stars we formed ratios of the extracted IRS spectra to the Kurucz model appropriate for the effective temperature and metallicity of each star. From the average of these ratios at each wavelength, we created a ``super-flat'' response curve for the SL1, LL2 and LL1 modules. Because of possible variability in the flat response in  the SL1 module, we used only the 8 other stars observed during the same instrument campaign (April 2004) as HD69830 to generate the super-flat for this module. For the LL modules we used data from all 35 stars. HD69830 was excluded from the creation of the super-flats.

\end{itemize}

The corrections derived from the comparison of extracted spectra and the photospheric models for the 3 modules are small:

\begin{itemize}

\item Shortward of 14~$\mu$m, the SL1 super-flat typically has values in the range 0.97-1.03, or, equivalently, deviations from unity response of $\sim$3\%; a few pixels have values deviating from unity by 10\%. The values at each pixel of the SL1 super-flat have a dispersion around the pixel response (averaged over all 8 stars) of $\sigma_{pop}\sim 2\%$ and $\sigma_{mean}\sim 0.4\%$. In the LL1 and LL2 modules, the typical super-flat values are in the range 0.98-1.02 with dispersions (averaged over all 35 stars) of $\sigma_{pop}\sim 1-2\%$ and $\sigma_{mean}\sim 0.2-0.3\%$. We divided the extracted spectra by the appropriate ``super-flat'' to remove any residual calibration variations. The dispersion in the super-flat ($\sigma_{pop}$) is an indication of the limiting systematic noise in removing the photospheric contribution to the signal from these stars.

\item It was necessary to adjust the absolute flux scale of all of the IRS/SSC spectra by a factor of 1.20$\pm$0.01 to fit the photospheric models obtained by fits to shorter wavelength (0.4-2.2~$\mu$m) data. With this adjustment the SL1 data are consistent with the IRAS 12~$\mu$m data and the LL2 data are consistent with the IRAS/MIPS data at 24~$\mu$m. The origin of this absolute calibration factor, e.g. pointing errors or an imperfect aperture correction, is being pursued but is not relevant to the present investigation.

\end{itemize}

\subsection{Results}

\begin{figure}
\includegraphics[width=4.7in, angle=-90]{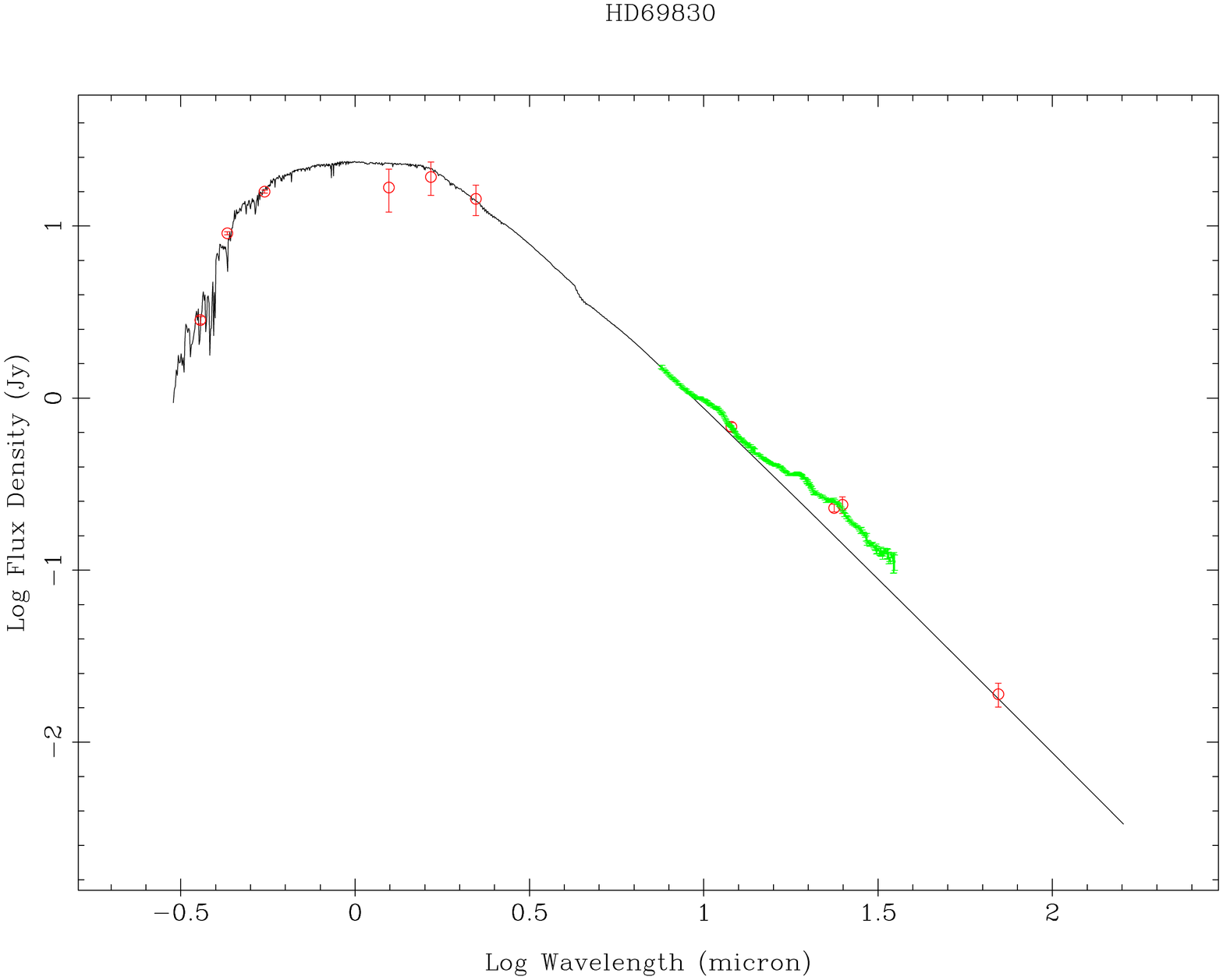}
\figcaption{{The Spectral Energy Distribution of HD69830 shows a Kurucz photospheric model (black) fitted to short wavelength data from Hipparcos and 2MASS (red). The 2MASS data (J, H, K$_s$) are quite uncertain due to the brightness of the star. Corrected spectra from the three IRS modules (green) are shown along with MIPS and IRAS photometry (red).}
\label{FullSpectrum}}
\end{figure}

\begin{figure}
\includegraphics[width=4.7in, angle=-90]{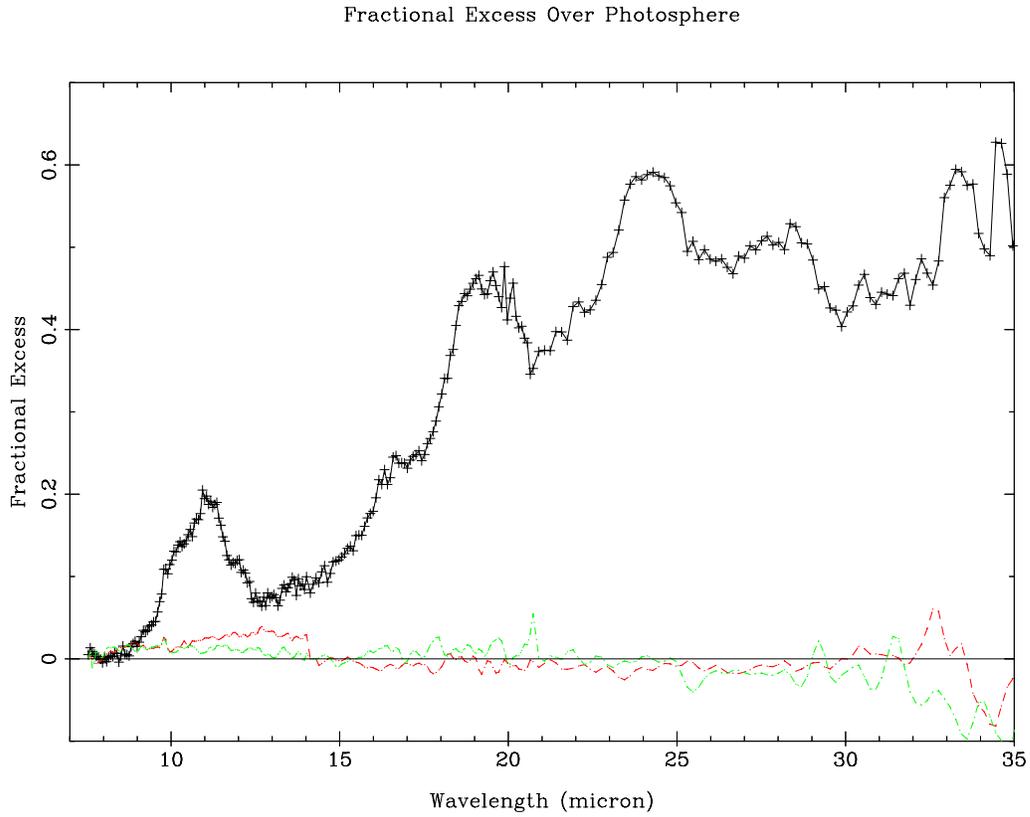}
\figcaption{After correction with the "super-flat" small deviations from a smooth photosphere are readily detectable in these IRS spectra. Shown are the fractional excesses, $(F_{\nu, Obs}-F_{\nu, Kurucz})/F_{\nu, Kurucz}$ for 2 stars with no evidence for excess at any wavelength, HD173667 (red) and HD142373 (green) and for HD69830 (black) which shows prominent excesses from 8 to 35~$\mu$m.
\label{residuals}}
\end{figure}

The overall Spectral Energy Distribution (SED) of HD 69830 including photometry (Table~\ref{phottable}) and the IRS spectra is shown in Figure~\ref{FullSpectrum}. A prominent excess is obvious in all three IRS modules reaching 60\% over the photospheric continuum at 35~$\mu$m. A number of broad spectral features, discussed below, are prominent in the data. Figure~\ref{residuals} compares the fractional deviation from a smooth photosphere for HD69830 with the spectra of two stars, HD173667 (AOR 4024320) and HD142373 (AOR 4023040), that show no excesses at either IRAS or MIPS wavelengths. As is clear from this figure, the size and spectral content of HD60930's excess are large compared to the uncertainties and will not be qualitatively altered by further refinements in the spectral extraction. Figure~\ref{residuals} can also be taken as a limit to the deviations which might be found toward any star, approximately 2-3\%. 

Included in Figure~\ref{FullSpectrum} and presented in Table~\ref{phottable} is MIPS photometry (AOR 4041728) as well as color-corrected IRAS data from the Faint Source Catalog \citep{moshir1990}. The MIPS data were reduced using the MIPS/GTO DAT software \citep{gordon2004} and show a prominent (50\%) excess at 24~$\mu$m, confirming the excess hinted at in the IRAS data at 25~$\mu$m \citep{mannings1998}. There is no evidence for extended emission in the 24~$\mu$m image. The consistency of the IRAS and Spitzer/MIPS data at 24 and 25~$\mu$m suggests that there has been no strong variation on the 20 year timescale.

As discussed elsewhere, HD69830 is the only star out of a sample of 84 showing a significant 24~$\mu$m excess \citep{beichman2005, bryden2005}. The uniqueness of HD69830 in this sample, 1 of 84 or 1.2\%, confirms the rarity of hot excesses around mature FGK stars first observed by IRAS \citep{fajardo2000} and ISO \citep{laureijs2002}. But HD69830 is even more exceptional in that its 24~$\mu$m excess {\it is not} accompanied by a longer wavelength excess as is normally the case for debris disk sources \citep{mannings1998, backman1998}. The measured value of the HD69830 excess at 70~$\mu$m is $1\pm3$ mJy compared to the predicted photospheric level of 18 mJy, or $F_{excess}/F_{star}<$50\% 
(3 times the 1 $\sigma$ statistical uncertainty combined in quadrature with a 20\% overall calibration uncertainty) at 70~$\mu$m. The lack of a 70~$\mu$m excess is evidence for the presence of small grains with low, far-infrared emissivity.

\begin{figure}
\includegraphics[width=5.5in]{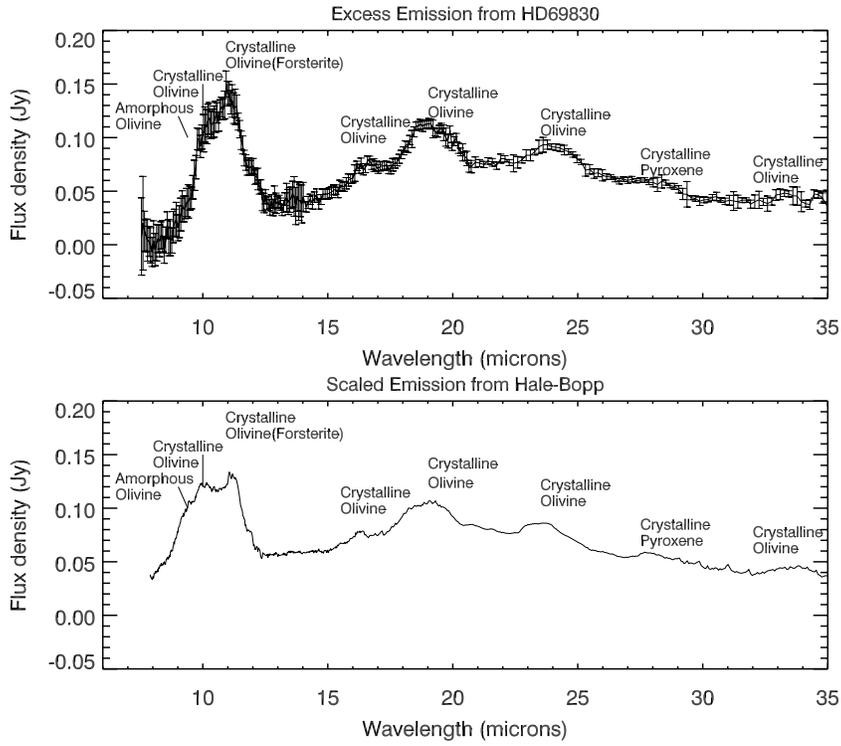}
\figcaption{{Top) The spectrum of the excess of HD69830; bottom) for comparison, the spectrum of the comet Hale-Bopp \citep{crovisier1996} normalized to 400K as described in the text. Identifications of some of the mineralogical features (Table~\ref{excesstable}) are indicated.}
\label{excess}}
\end{figure}

The upper panel of Figure~\ref{excess} shows the IRS data with the photospheric component subtracted to reveal the details of excess. Uncertainties at each wavelength were derived using the largest of two terms: a) the quoted errors in the individual SPICE extractions; and b) the difference between the corrected spectra obtained at the two different nod positions. To give the total uncertainty, the preceding value is combined in quadrature with a systematic term equal to 1\% of the photospheric emission to account for errors in the subtraction of the photosphere. The first two terms reflect statistical uncertainties in the spectrum and become worse at the longest wavelengths where the signal to noise is poorest. The third term, being proportional to the stellar brightness, decreases in importance at longer wavelengths.

\section{The Excess Toward HD69830}

The lower panel of Figure~\ref{excess} shows a spectrum obtained by the Infrared Space Observatory (ISO) of the comet Hale-Bopp \citep{crovisier1996} and scaled as described in $\S$4.2 to match the overall slope of the Spitzer data. The detailed similarity between the two spectra is remarkable and will be used as a guide to the identification of features in the spectrum of HD69830.

\subsection{Features in the Spectrum of the excess}

Table~\ref{excesstable} identifies some of the features in the excess. Most prominent are broad features at 9.4-12~$\mu$m, 16~$\mu$m, 19.3~$\mu$m and 23.8~$\mu$m. Other features of moderate strength include a plateau at 10~$\mu$m, a plateau at 27~$\mu$m, and a weak feature of low signal-to-noise ratio around 33-34~$\mu$m. As noted in Table~\ref{excesstable} and Figure~\ref{excess}, small Mg-rich crystalline silicate grains\footnote{As discussed in Wooden et al. (1999 and 2000) olivines are minerals with [Mg$_y$Fe$_{1-y}$]$_2$SiO$_4$ and $0 \leq y\leq 1$; forsterite is an Mg-pure olivine with y=0. Pyroxenes are minerals with [Mg$_x$Fe$_{1-x}$]SiO$_3$ and $0 \leq x\leq 1$; enstatite is a Mg-pure pyroxene with x=0.}, e.g. enstatite and forsterite, have spectral features that match those in the spectrum remarkably well \citep{jager1998, koike1993}. These are seen in the ISO and ground based spectra of Hale-Bopp \citep{crovisier1996, crovisier1997, wooden1999, wooden2000}. While not dominant in the spectrum of the zodiacal cloud, these crystalline silicates are found in interplanetary dust particles as well as in meteorite inclusions \citep{yoneda1993, bradley2003}.

A detailed examination of the spectrum reveals that forsterite matches the 9-11~$\mu$m feature, while enstatite more closely matches the ratio of the 19~$\mu$m and 23.8~$\mu$m features (Henning 1998). Features due to crystalline olivines are far more prominent than those due to pyroxenes, although an inflection point around 27~$\mu$m might be due to crystalline pyroxene. The 9.3~$\mu$m feature characteristic of small, amorphous silicate grains that is prominent in Hale-Bopp and seen weakly in our own zodiacal cloud \citep{reach2003} is not seen at all in HD69830. Since the emission between the features does not drop to zero, there must be some source of continuum opacity, possibly larger grains, in both HD69830 and Hale-Bopp. In summary, we attribute the excess toward HD69830 to a mixture of grains similar to those seen in Hale-Bopp and very different from those responsible for the relatively featureless spectra of typical debris disks. 

\subsection{Model for the Excess Emission}

The simplest model for the emission (Figure~\ref{excess}) scales the ISO spectrum of Hale-Bopp by the ratio of two blackbodies:

$$F_\nu(HD69830) \propto F_\nu (Hale\, Bopp) \times B_\nu(400 K) / B_\nu(T(Hale\, Bopp)) \eqno{(1)} $$

\noindent where $B_\nu$(T) is the Planck function and T(Hale-Bopp)=207 K is the equilibrium temperature of a small crystalline silicate grain 2.9 AU from the Sun. The value of 400 K is fitted by eye to provide a reasonable fit to the HD69830 excess and corresponds to an equilibrium distance of $\sim$0.5 AU for small grains. Although the single temperature scaling gives a remarkably good representation of the data, we next develop a more realistic model that constrains the location, size, and composition of the emitting material. 

The presence of pronounced spectral features on top of a weak continuum suggests optically thin emission dominated by dust grains comparable to or smaller than the wavelength: $ a\leq \lambda_c / 2\pi \sim 1\, \mu$m where $\lambda_c \sim 8\, \mu$m is the shortest wavelength showing spectral features. The close analogy to Hale-Bopp requires the existence of grains as small as 0.25~$\mu$m \citep{wooden1999, wooden2000}. While fluffy, low density grains with a size around 1~$\mu$m have been suggested to fit the spectrum of Hale-Bopp \citep{li1998}, for this preliminary analysis we adopt the following empirical model:

\begin{itemize}

\item Synthesize empirical absorption cross sections, $Q_{abs}$, derived
from the scaled Hale-Bopp observations \citep{wooden1999, wooden2000}, $Q_{abs} \propto F_\nu / B_\nu(T_{HB}), $ with $T_{HB}=207$ K and scaled to match laboratory measurements of crystalline olivine at 11.3~$\mu$m \citep{mukai1990}. 

\item Subtract optical constants for 0.25~$\mu$m amorphous silicate grains \citep{draine1984, weingartner2001} from the optical constants derived from Hale-Bopp since the 9.3~$\mu$m feature produced by this type of grain is not seen in HD69830. Examination of Hale-Bopp emission models, e.g. Figure 6 in Wooden et~al. (1999) suggests that amorphous grains are at least a factor of five less abundant in HD69830 than in Hale-Bopp.

\item Synthesize empirical absorption cross sections at 70 $\mu$m as described above using the 70 $\mu$m data for Hale-Bopp from Gr\"un et~al. (2001). The assumption that all the emission comes from grains following the same equilibrium temperature law is naive but gives a self-consistent estimate of the 70 $\mu$m emission. 

\item Assuming a flat disk geometry for convenience, calculate the temperature gradient through radiative equilibrium of 0.25 $\mu$m grains orbiting a K0 star and integrate the surface brightness profile, $I_\nu(r)=\tau(r)B_\nu(T(r))$ over a region extending from the grain sublimation distance (at which T$\sim$1500 K and corresponding to $r\sim 0.06$ AU) out to an outer radius $r_{max}$.

\item Let the radial dependence of the optical depth follow a power law distribution, $\tau(r) \propto r^\alpha$. For comparison, $\alpha=-0.4$ in our solar system.

\end{itemize}

\begin{figure} 
\includegraphics[width=4.7in]{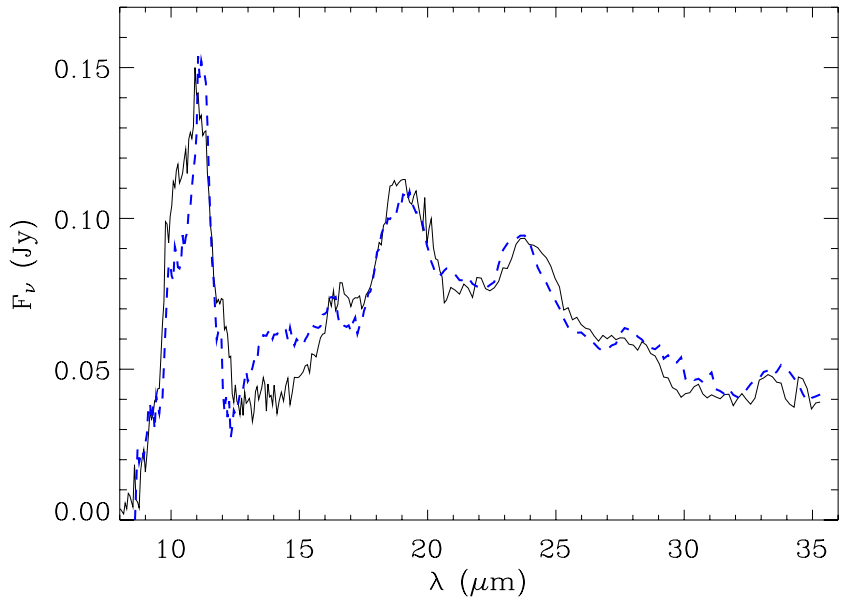}
\figcaption{{As described in the text, a simple model using crystalline
silicate grains in a disk extending to $\sim1$ AU (dashed blue line) is fitted to the IRS spectrum of the excess (solid line).}
\label{model}}
\end{figure}

A $\chi^2$ search was made through a broad range of values for the free parameters in the model: the surface area of the small grains derived from the Hale-Bopp spectrum, the power law index of the surface density profile, and $r_{max}$. The parameters of the best fit model are given in Table~\ref{modeltable}. As shown in Figure~\ref{model}, the model reproduces many, but not all, of the details in the IRS spectrum. Some of problems include a shoulder at 10 $\mu$m that cannot be fit without throwing off the fit at 11 $\mu$m; an excess between 13-15 $\mu$m not seen in Hale-Bopp (the feature does not correspond to known Mg- or Mg/Fe-rich silicates, but resembles unidentified features in the spectrum of R Cas \citep{jager2003}); subtle variations around 19~$\mu$m and 24~$\mu$m where \ 4 or 5 separate bands of forsterite and enstatite make up a broad feature sensitive to small composition and/or temperature differences \citep{bowey2001}. Despite these small but significant differences in detail, the Hale-Bopp based model provides an excellent analogue for the HD69830 debris disk.

In the final fit (Table~\ref{modeltable} and Figure~\ref{model}) we constrained the power law index of the radial surface density power law to have a value of $-0.4$ since varying this parameter away from -0.4 did not significantly improve the overall results. Models with steeper power law indices and $r_{max} \ge 5$ AU were consistent with the limit to the $70~\mu$m excess, but had poor fits in the $7-35~\mu$m region with $\chi^2 \ge 10$. Models with $\alpha =-0.4$ and $r_{max}>1$ AU also had poor $\chi^2$ values and were inconsistent with the $70~\mu$m limit. Although grain size was not an explicit free parameter in the model because we adopted opacities directly from the Hale-Bopp spectrum, we found that grains larger than about $5~\mu$m were inconsistent with the limits on the $70~\mu$m excess for any reasonable combination of model parameters.

The fractional luminosity in the excess, integrated between 7 and 35~$\mu$m is $ L_{dust}/ L_* =2\times 10^{-4}$, making this system a factor of two brighter than Fomalhaut ($\alpha$ PsA; Backman and Paresce 1993; Stapelfeldt et al. 2004) in terms of fractional dust luminosity and more than 1,000 times brighter than our own zodiacal cloud. The surface area of emitting material in the HD69830 cloud is $1.3\times10^3$ times that of our own zodiacal cloud, $\sim 2 \times 10^{20}$ cm$^2$ within 2-3 AU \citep{backman1998}. In terms of a fractional surface area of emitting material, our solar zodiacal cloud has $\Sigma_{frac}\sim 1\times 10{-7}$ which compares to the  fractional surface area of 0.25 $\mu$m grains within 1 AU of $\Sigma_{frac}\sim  3\times 10^{-4}$ for  HD69830.

If the emitting grains all have radius $a=0.25~\mu$m with density $\rho= 3$ g cm$^{-3}$ then the mass of the emitting material (Table~\ref{modeltable}) would be ${4 \over 3} \, \Sigma \, \rho \, a= 2.7\times 10^{19}$ g or $4.6\times10^{-9}$ M$_\oplus$ where $\Sigma=N_{tot}\pi a^2$ is the emitting surface area of $N_{tot}$ grains. However, since HD69830 is probably surrounded by a broad distribution of grain sizes, the total mass could be much larger. For a grain size distribution $n(a)=n_0 ( a / a_0 )^{-3.5}$, corresponding to predictions for a collisional cascade \citep{dohnanyi1969}, the ratio of total mass, $M$, to surface area, $\Sigma$, is $ (M / \Sigma)\sim { 4 \over 3} \rho \sqrt{a_{min} a_{max}}$ where $a_{min}$ is the minimum grain size and $a_{max}$ is the maximum grain size. For $a_{min}=0.25~\mu$m and $a_{max}=10$ km, the total mass of material associated with the disk emission (Table~\ref{modeltable}) is $5.4\times 10^{24}$ g or $0.9 \times10^{-3}$ M$_\oplus$; given the strong dependence on the unknown grain size distribution, this extrapolation is obviously very uncertain. This number should be compared to the cumulative mass of asteroids up to 10 km in our solar system, $0.014\times10^{-3}$ M$_\oplus$ \citep{bidstrup2004, krasinsky2002}, indicating a factor of 64 times more mass in a putative ``asteroid" belt around HD 69830 than in our own asteroid belt (up to 10 km size). We analyze this estimate in the context of grain production and loss mechanisms in $\S4.4$.

\subsection{Limits on a Kuiper Belt}

One of the most striking aspects of HD69830 is the lack of a long-wavelength excess to accompany the 8-35 $\mu$m excess. Treating the upper limit at 70~$\mu$m just like the upper limits toward other main sequence stars observed by MIPS \citep{beichman2005, bryden2005} yields an estimate for the amount of cold dust. The peak of the cold dust emission comes at 70~$\mu$m for material at a temperature of 50 K located at $\sim$ 50 AU from the star (for small grains; perhaps a factor of 2 closer in for larger grains). In this case, $L_{dust}/ L_* < 10^{-5}\, \, (5600\, K / T_*)^3 \, \, (F_{dust}/ F_*) \sim 5\times 10^{-6} \, \, \, (3\, \sigma)$. This is only 5 times larger than the upper end of the range of values, $10^{-6}$ to $10^{-7}$, predicted for the Kuiper Belt region beyond 30 AU in our solar system \citep{stern1996} and less than the IRAS constraint, $<10^{-5}$ \citep{aumann1990}. The HD69830 limit on cold material is uncertain as it is based on only a single wavelength; observations in the submillimeter would further constrain the potential reservoir of distant cold material \citep{greaves2004}. 

\subsection{Origin of the Emitting Grains}

Nothing in the properties of HD69830 suggests that this star is anything other than a mature, main sequence dwarf. Thus, there is no reason to attribute its excess to dust associated with the weak mass loss often seen toward M and K giants \citep{beichman1990}. Spectral features in the excess emission around such stars have been identified with small crystalline grains condensing in the quiescent, outflowing gas \citep{henning1998, dijkstra2003}, but since HD69830 is a quintessential dwarf star, we do not consider this possibility further. In the absence of any purely stellar phenomenon, we attribute the excess to orbiting dust derived from the solid material left over from the formation of the star itself, i.e. a debris disk. We now investigate the source of this dust in more detail. 

\subsubsection{Timescales}

The rarity of main sequence stars with hot excesses and without an associated colder debris cloud ($\leq$1\%) suggests that the conditions for this excess are found around fewer than 1 star in 100 or that it is a transitory (or low duty cycle) event lasting no more than 1\% of the 2 Gyr lifetime of HD69830. 

What are the natural timescales of the debris clouds in HD69830, and, for comparison, in our solar system? A debris cloud continually loses material due to collisional grinding down of grains to a size small enough that radiation pressure and/or Poynting Robertson (PR) drag ultimately remove them from the system. The relevant timescales \citep{burns1979, bp1993} are $$\tau_{coll}= { t_{orbit} \over 8\Sigma_{frac}} = { r_{AU}^{1.5} \over 8\Sigma_{frac} {M_*}^{0.5} } \, {\rm yr} \eqno{(2)} $$ where $\Sigma_{frac}$ is the {\it fractional} surface area of the dust \citep{bp1993} and the PR drag time \citep{bp1993} $$\tau_{PR}=700 \, a_{grain} \, \rho \, r_{AU}^2 \, {L_\odot \over L_*} \, \, {\rm yr} \eqno{(3)} $$ where $\rho$ (g cm$^{-3}$) is the particle density and the grain size is in $\mu$m. 

Table~\ref{timescales} compares relevant parameters for our solar system's zodiacal cloud and that of HD69830. In our solar system the critical (shortest) timescale is the PR time, $\tau_{PR}=2-6 \times 10^5$ yr for 10-30~$\mu$m grains \citep{fixsen2002, reach2003}. In HD69830 the collisional timescale is 400 yr and the PR timescale about 700 yr for 0.25~$\mu$m grains at 1 AU. The grains will grind themselves down through collisions until they are small enough to be lost through PR drag, i.e. until $\tau_{PR} \sim \tau_{coll}$.  Following Sheret et al. (2004; see their Figure 3) we note that radiation blowout does not operate efficiently for low luminosity stars like HD69830 and $\epsilon$ Eri, particularly for non-idealized grains that are either porous or non-spherical \citep{li1998}.  Collisions followed by loss via PR drag are likely to dominate the grain distribution around HD69830. Although the uncertainties on the HD69830 timescales are large (at least a factor of two due to unknown grain size and orbital location), it is clear that small grains must be replenished on roughly  the $<$1,000 yr timescale or the emission associated them will disappear.

\subsubsection{An Enhanced Zodiacal Dust Cloud}

The probable source for the small grains seen toward HD69830 is a debris cloud located within 1 AU of the star that is replenished on timescales of $10^3$ yr. For comparison, our zodiacal cloud consists of relatively large grains (10-100~$\mu$m) extending to 3 AU and maintained in rough equilibrium, with occasional upward spikes, through asteroid collisions every few million years \citep{fixsen2002, dermott2002a, dermott2002b}. 

The mass in 0.25~$\mu$m dust grains orbiting HD 69830 ($\S$4.2), $4.6\times 10^{-9}$ M$_\oplus$, is destroyed in $\tau_{coll} \sim \tau_{PR}\sim 0.5-1\times 10^3$ yr for a mass loss rate of $0.5-1\times 10^{-11}$ M$_\oplus$ yr$^{-1}$. This loss mechanism represents the end point for the small grain debris generated by the collisional cascade of larger bodies into smaller ones. If this mechanism operates continuously over 2 Gyr, then the total mass of solid material lost to the system over 2 Gyr would be $ > 10^{-2}$ M$_\oplus$, or more than 500 times the amount of material in own asteroid belt. Note that the estimate of solid material lost over time is also considerably greater than the present day mass estimate, $1 {\rm \ to} 3\times 10^{-3}$  M$_\oplus$ inferred from integrating an $a^{-3.5}$ distribution up to either 10 or  100 km bodies ($\S$4.2). Either we are seeing the decaying remnant of a more massive disk or the mass loss is episodic and the average mass loss rate is not as great as present day observations would suggest. 

Another way to estimate the amount of material present is to examine the dust production rate. If, as noted above (Sheret et al. 2004), radiation blowout is not effective around HD69830, then PR drag is the dominant loss for dust in  HD69830 and the number of emitting grains in the disk, $n_{gr}$, is simply proportional to the square of the number of large bodies (comets or asteroids), $N_c$, that initiate the collisional cascade leading to these small grains: $n_{gr} \propto N_c^2$ \citep{dominik2004}. Presently, our zodiacal cloud has a fractional surface density around $10^{-7}$, but immediately after a major collision, the cloud would have a larger population of small grains given by the n(a)$\propto a^{-3.5}$ law appropriate to a collisional cascade \citep{dohnanyi1969}. The surface area of emitting material and the resultant optical depth of such a cloud is $\propto \int a^2 \, a^{-3.5}\, da \propto a^{-0.5}$. Extending the small grain cutoff from its present 10~$\mu$m to 0.25~$\mu$m would temporally increase the emitting surface area by a factor of at least a factor  $\sim 6$ and probably more since a collision would increase the amount of dust of  all sizes. If we compare the cloud around HD69830 with our solar system (enhanced to include short-lived 0.25~$\mu$m grains for this calculation), we need $\Sigma_{frac}$(HD69830) / $\Sigma_{frac}$(Solar System) = $3\times 10^{-4}/6\times 10^{-7} = 500$ times more small grains orbiting HD69830 than our zodiacal cloud would have immediately after a major collision.

Since, as argued above, $\Sigma_{frac}\propto N_c^2$ \citep{dominik2004}, the number of large bodies required to replenish the dust would be $N_c \propto \sqrt{500}=22$ more than in our solar system. By this reasoning the total mass of a putative asteroid belt would be 0.3 $\times 10^{-3}$ M$_\oplus$ which compares favorably with the value of 1 $\times 10^{-3}$ M$_\oplus$ derived from integrating an $a^{-3.5}$ size distribution up to 10 km. Although both mass estimates would increase if the upper size mass limit were increased to 100 km, the values are considerably smaller than the $10^{-2}$ M$_\oplus$ estimated by integrating the instantaneous mass loss rate over the maximum 10$^7$ yr  lifetime of the excess. The simplest hypothesis is that the numbers are consistent within the uncertainties (particularly since the derived masses depend on the unconstrained upper end of the integral over the particle size distribution, 10 $\sim$ 100 km). If the difference between the instantaneous and long term mass rates is significant, then we could be seeing a transient peak in surface area, possibly  from a very recent collision  ($<\tau_{PR}\sim10^3$) yr.  

The exact details of an asteroid belt model will have to be established by a combination of improved modeling to constrain grain properties and location more precisely, as well as more realistic modeling of the dynamics. In any event,  HD69830 could have  an asteroid system with much higher density located much closer to the star than in our solar system. Pursuing the analogy with our solar system suggests that an as-yet-unseen planet located outside r$_{max}$ could truncate the disk and confine most of the dust to radii interior to the planet's orbit \citep{menou2003}. Such a planet could also enhance the lifetime of dust by trapping it into mean motion resonances \citep{kuchner2003}.

\subsubsection{A Swarm of Comets as a Source of Small Grains}

An alternative to  emission from small grains in an enhanced asteroid belt is emission from cometary ejecta. Levison et al. (2001) discuss possible mechanisms for stirring up the Kuiper Belt including scattering of material into the inner solar system by the late formation of distant planets or the destabilization of Kuiper Belt objects due to sudden changes in the orbital parameters of one or more giant planets. Alternatively, some perturbing event might send a single large object located far from the star, i.e. a 1,000 km sized object like Sedna \citep{brown2003}, into the inner HD69830 system where it becomes a ``super-comet''. We examine these two possibilities for HD69830 below.

What would it take to reproduce the emission from HD69830 using cometary debris? Consider the F$_\nu(11.3\, \mu {\rm m})= 7300$ Jy flux density of Hale-Bopp measured in a 9.3\arcsec\ beam when Hale-Bopp was located at $r_h=1.1$ AU from the Sun and $\Delta_{HB} =$ 1.6 AU from the Earth \citep{williams1997}. If we scale this to Hale-Bopp's perihelion values of $r_h=0.9$ AU and $\Delta_{HB} =$ 1.5 AU according to $r_h^{-4} \Delta^{-1}$ \citep{gehrz1992, wooden1999} and to a radius of 20\arcsec\ \citep{hayward2000}, then the total 11.3 $\mu$m emission of this object would be $\sim 6 \times 10^4$ Jy. Placed at the $\Delta(HD69830)= 12.6$ pc distance of HD69830, it would take 

$$N_{HB}={ {F_\nu(HD69830)} \over {F_\nu(HB)} } \left ( { {\Delta(HD69830)} \over {\Delta(HB)} } \right ) ^2 \sim 6 \times10^6 \eqno{(4)} $$ comets identical to Hale-Bopp as it appeared at 0.9 AU to reproduce the 0.013 Jy flux density measured by Spitzer at 11.3 $\mu$m. Reducing the periastron distance to 0.5 AU consistent with the $\sim 400$ K temperature determined in the modeling would increase the brightness of the cometary material and thus decrease their required number to $\sim 5\times 10^5$ Hale-Bopps.

Consider a steady stream of comets on $Period=10^3$ yr orbits arriving at the inner HD69830 system at a rate of $\eta$ comets yr$^{-1}$. If each comet is bright while it is within 3 AU for a time $t_{3 AU} \sim 1$ yr and makes $N_{Pass} \sim 100$ active passes through the inner HD69830 system \citep{grun2001, wooden2000}, then in steady state $N_{HB}=\eta t_{3AU}$ comets contribute to the observed excess at any one time; $N_{active}=\eta Period $ contribute to the excess whenever they come close to the star; and $N_{total}=\eta \, Lifetime/N_{Pass}$ comets contribute to the excess during the putative $10^7$ year lifetime of the excess. If, as argued above, we need $5\times10^5$ Hale-Bopps at any instant, then $\eta=5\times 10^5$ comets yr$^{-1}$. With these numbers, $N_{active}=5\times 10^8$ and $N_{total}=5\times 10^{10}$ over the duration of this excess. With a total mass for Hale-Bopp of $ \sim 1 \times 10^{20}$ g, assuming a density of 1.5 g cm$^{-3}$ for a mixture of grain and ice and a radius of $\sim$ 25 km \citep{weaver1997}, then the total mass in comets entering the inner HD69830 system during this $10^7$ yr period would be $\sim 900 M_\oplus$, unreasonably large for any residual Kuiper Belt. Even if more normal, smaller comets were adopted, e.g. a factor of 10 smaller than Hale Bopp, the total mass would still require $\sim 90 M_\oplus$ in comets which still seems large given the 70 $\mu$m limts on a distant Kuiper Belt.  While the uncertainty in this simple model is large, the total mass of comets required to maintain the hot material argues against the cometary hypothesis.

\subsubsection{A ``Super-Comet'' as a Source of Small Grains}

An alternative to a swarm of comets is a single giant comet scattered into the inner HD69830 system, perhaps an object like Sedna with a radius of 1,000 km and composed of ice and rock. Such an object would have a surface area $(1000/25)^2=1600$ times that of Hale-Bopp and in an low eccentricity, 0.5 AU orbit would evolve gas and small dust grains at a rate $dM(t) / dt$=1600 times greater than Hale-Bopp. In an idealized picture, this material would build up over $\tau_{PR}$, would last until the entire comet evaporated in a time, $\tau_{evap}$, and then decline in $\tau_{PR}$. During $\tau_{evap}$ the average amount of material would be equivalent to $ dM(t) / dt \times \tau_{PR}\sim 1.6\times10^6$ Hale-Bopps which agrees reasonably well with the amount of emitting material estimated in $\S 4.4.3$.

We can estimate the evaporation time by noting that the mass loss rate is proportional to the surface area of the comet: $dM(t) / dt=4 \pi \Phi R(t)^2$ where $\Phi$ is the flux of material emitted in g cm$^{-2}$ s$^{-1}$ and R(t) is the radius as a function of time. Then, with $R(t)= ( 3 M(t) / 4 \pi\rho ) ^{1/3}$ and $M(0)= 4/3 \pi \rho R(0)^3$, we get $$M(t)= { {4 \pi (R(0) \rho -\Phi t)^3} \over {3 \rho^2}} \eqno{(5)} $$ and $\tau_{evap}=R(0) \rho / \Phi$. At perihelion, Hale-Bopp gave up some 200 metric tons sec$^{-1}$ in gas and small grains \citep{lisse1997} and possibly as much as 2,000 metric tons sec$^{-1}$ in larger grains detected at submillimeter wavelengths \citep{jewitt1999}. For Hale-Bopp with an assumed 25 km radius and a mass loss rate of $200$ tons sec$^{-1}$ for small grains, $\Phi=76$ g cm$^{-2}$ s$^{-1}$. An escape velocity from a Sedna-like object is about 0.5 km s$^{-1}$, half that of Pluto's, and comparable to the thermal velocity of H$_2$O at 300 K. Small grains entrained with the gas could escape the nucleus while larger grains would probably fall back under the self-gravity of the object. Therefore, we adopt the lower, small-grain mass loss estimate for this putative super-comet. It is worth noting  that this grain segregation mechanism would account naturally for the lack of large grains with their associated long wavelength emission. 

A 1,000 km sized comet trapped in a low eccentricity orbit (making $\Phi$ roughly constant) would have an evaporation lifetime of $\tau_{evap}=$ 2 Myr and could provide material over a long enough time to have a reasonable chance of being observed by Spitzer. Effects of self-gravity and ice-rock differentiation would reduce the flux of evolved material, but would extend the duration of the outburst. This scenario would be less attractive if the super-comet were on a highly eccentric orbit, spending much of its time away from the star. The requirement for the super-comet to travel in a low eccentricity orbit close to the star suggests the presence of a planet located close to the star to effect its capture. 

Finally, we note that the rupture of a single Sedna-like object in the inner HD69830 system could release a swarm of $10^5-10^6$ smaller, 10-25 km bodies which in turn could produce a large number of small grains as envisioned in the preceding section. Since the small grains would last only $\tau_{PR}$, a significant fraction of the swarm would have to survive for a few million years to replenish the small grains and thus to make the outburst last long enough for us to have a chance of seeing it. 

In either case, the Sedna-like object could have been dislodged from a stable, more distant orbit by a planet in the HD69830 system or by the effects of a passing star. The Sedna-like object might even have been captured from the passing star itself \citep{kenyon2004b}, although the space motion of HD69830 \citep{eggen1998} indicates no nearby approaches to any known star within the past 20 Myr (V. Makarov, private communication, 2005).

\subsubsection{An Observational Test for the Origin of the HD69830 Excess} 

There is a powerful observational test that could decide between the two models. The cometary scenario would bring a great deal of volatile material into the inner HD69830 system. In Hale-Bopp, species such as CO, H$_2$O, HCN, and oxygen were detected at wavelengths from the millimeter to the ultraviolet. Sensitive spectroscopy of HD69830 might reveal evidence for volatiles seen either in emission or in absorption against the star, similar to what has been detected toward $\beta$ Pictoris in the visible \citep{lagrange1992, petterson1999} and in the UV \citep{vidal1994, roberge2000}, and toward HD100546 \citep{vieira1999, bouwman2004}. While examination of the IUE archive spectrum of HD69830 \citep{hufnagel1994} shows no features attributable to gas in a shell around the star as is seen in $\beta$ Pictoris, e.g. excess Mg II absorption at 279.5 and 280 nm, more careful searches for volatile material are clearly warranted. Detection of such volatiles would strongly support the cometary hypothesis and represent a milestone in our understanding of how organic molecules are transported from the cold outer solar system into the region of the terrestrial planets.

\section{The Detectability of Planets in the HD69830 System}

Finally, we note that the intense diffuse emission toward this star makes it unlikely that future astronomical telescopes will ever detect directly the light from terrestrial planets in the HD69830 system. For telescopes like the Terrestrial Planet Finder \citep{beichman2003}, the photon noise at either visible or mid-IR wavelengths associated with a target's exo-zodiacal cloud begins to dominate other noise sources at levels exceeding 10 times the brightness of our own zodiacal cloud. Unfortunately, the HD69830 cloud, being more than 1,000 times brighter than our cloud, will hide planets like our own from view. 

From the standpoint of TPF it is useful to note that the limiting IRS sensitivity described herein will allow us to detect or set limits to small grains located in the ``habitable zones'' at least a factor of 10 fainter than that around HD69830. A focused program of IRS observations should reach $\sim 100$ times the level of our zodiacal cloud and thus provide a good initial filter for TPF target selection. 

\section{Conclusion}

We have identified an excess of emission between 8 and 35~$\mu$m toward the nearby K0 V star HD69830. The IRS spectrum of the excess reveals prominent features due to crystalline silicates such as forsterite, which must be present in small, sub-micron dust grains for the features to be so prominent. The lack of longer wavelength emission and the ratios of various spectral features suggest that most of this material has a temperature in excess of 400 K and must be located within 1 AU of the star. An excess of this type (strong at 25 $\mu$m, but weak at 70 $\mu$m) is unique among the sample of 84 main sequence stars examined by Beichman et al. (2005) and in other IRAS/ISO samples, implying a short lived event or a low duty cycle process. 

Estimates of the mass in an enhanced asteroid belt needed to replenish the small dust grains seen by Spitzer are quite uncertain: a) 64 times the amount in our solar system based on an extrapolation of the emitting surface area of small grains up to 10 km bodies; or b) 22 times the amount in our solar system based on dust production in asteroid-asteroid collisions. If the mass loss rate inferred from the present day excess were constant over 2 Gyr, the system would have lost at least 500 times the amount of asteroidal material in our solar system. It is more likely that the excess is sporadic and the total amount of lost ``asteroidal" material is smaller than this value.

Alternatively, small grains could arise from a single, Sedna-sized ``super-comet'' recently captured in the inner HD69830 system and ``evaporating" over a few million years or breaking apart into smaller objects. This super-comet would have to be trapped in a low eccentricity orbit around 1 AU from the star. Spectroscopic follow-up to search for volatiles associated with comets would help distinguish between these two hypotheses.

\section{Acknowledgments}

We would like to thank Geoff Marcy, Debra Fischer, and John Stauffer for valuable discussions on the properties of HD69830. Sergio Fajardo-Acosta provided programs we used for accessing the Kurucz models. We acknowledge useful discussions with Doug Lin on the relative merits of asteroid and Kuiper Belts as a source of the excess, Valeri Markarov on the recent trajectory of HD69830, and Bill Reach on the physical properties of comet trails. 

This research made use of the IRAS, 2MASS, and Hipparcos Catalogs, as well as the SIMBAD database and the VizieR tool operated by CDS, Strasbourg, France. 

The {\it Spitzer Space Telescope} is operated by the Jet Propulsion Laboratory, California Institute of Technology, under NASA contract 1407.  Development of MIPS was funded by NASA through the Jet Propulsion Laboratory, subcontract 960785. Some of the research described in this publication was carried out at the Jet Propulsion Laboratory, California Institute of Technology, under a contract with the National Aeronautics and Space Administration. 

Finally, we remember with great sadness the efforts of NRC postdoctoral fellow Elizabeth Holmes who worked intensively on this project before her untimely death in March 2004.




\clearpage

\clearpage

\clearpage


\clearpage





\clearpage




\begin{table}
\caption{Infrared Photometry of HD69830 \label{phottable}}
\begin{tabular}{lcccc} \\ 
Facility & $\lambda (\mu$m)& $F_\nu$(Jy)&Photosphere(Jy)$^1$ &Significance$^2$, $\chi$\\ 
\tableline 
IRAS&12.0&0.68 $\pm$0.041&0.61 &1.7\\ 
MIPS&23.7&0.23 $\pm$0.012&0.16 &6.1\\
IRAS&25.0&0.24 $\pm$0.026&0.14 &3.8\\
MIPS&70.0&0.019$\pm$0.003&0.018&0.4\\ 
\end{tabular}
\tablecomments{$^1$From Kurucz model fitted to short wavelength data}
\tablecomments{$^2\chi$ =(Observed- Kurucz)/$\sqrt{\rm noise^2+(model \, uncertainty)^2}$.}
\end{table}

\clearpage

\begin{table}
\caption{Features in Excess Compared with Identifications in Comet Hale-Bopp$^{1, 2}$ \label{excesstable}}
\begin{tabular}{lrr} \\
$\lambda (\mu$m) &Comment & Identification\\
 \tableline 
11 (9.4-12.0 ) &Very strong emission &Crystalline Forsterite\\
10.2 &Shoulder within 11 um feature& Crystalline olivine \\
 11.0-11.3&Peak within 11 um feature& Crystalline olivine \\
16.7 (16.2-17.5) &Broad feature on shoulder of 19 $\mu$m feature& Crystalline olivine \\
 19.3 (17.7- 20.7)&Very broad emission feature & Crystalline olivine\\
 23.8 (22.7-24.7)&Prominent, broad feature& Crystalline olivine\\
 27.8& Weak feature or plateau & Crystalline pyroxene (Enstatite)\\
 33 -35 &Weak feature & Crystalline olivine, low SNR\\
\end{tabular}
\tablecomments{$^{1, 2}$ Crovisier et al. 1996, 1997; Wooden
et al. 2000; Henning 1998.} 
\end{table}
\clearpage

\begin{table} 
\caption{Properties of Best Fit Model \label{modeltable}} 
\begin{tabular}{lr} \\
 &Model \\
 \tableline 
{\it Best Fit Model Parameters} &\\
Density power law index, $\alpha$ & -0.4 \\
\ \ \ \ (constrained) & \\
r$_{max}$ (AU) & 1.0 \\
Surface area of Hale-Bopp like grains$^1$ (cm$^{2}$) & $2.7\times10^{23}$ \\
Reduced $\chi^2$ (250 d.o.f) & 3.8 \\
\hline
{\it Derived Parameters} &\\
Predicted 70 $\mu$m flux density (mJy) & 0.75 \\
Dust Mass in 0.25 $\mu$m grains (M$_\oplus$) & $1.0\times10^{-9}$ \\
\end{tabular}
\tablecomments{
$^1$Assumes optical constants derived from scaled Hale-Bopp spectrum with
amorphous silicates subtracted to remove the 9.3~$\mu$m feature seen in Hale Bopp, but not in HD 69830.} 
\end{table}

\begin{table} 
\caption{ Comparison of Solar System and HD69830 Model \label{timescales}} 
\begin{tabular}{lrr} \\
Parameter &Solar System&HD69830 \\
\tableline 
{\it System Parameters (see text)} & & \\
Orbital Distance (AU) & 3 & 1\\
Stellar Mass (M$_\odot$)& 1 & 0.8\\
Stellar Luminosity (L$_\odot)$&1 & 0.45\\
Grain Size ($\mu$m) & 10-30& 0.25\\
Fractional Dust Surface Density & $10^{-7}$ ($<$3 AU)& $3\times 10^{-4}$ ($<$1 AU) \\
\ \ \ (enhanced after collision, see text) & $6\times 10^{-7}$ & \\
Fractional Dust Surface Density$^1$ ($>$30 AU)& $10^{-6}$ -$10^{-7}$ (estimate) & $<6\times 10^{-6}$\\
\hline
{\it Timescales for Dust Destruction ( 1 AU) } & & \\
$\tau_{coll}$ (yr) & 1$\times10^6$ & 400\\
$\tau_{PR}$ (yr) & 2-6$\times10^5$ & 700\\
\hline
{\it Asteroid Belt Mass  (a$_{max}=10$ km; M$_\oplus$)} & & \\
Our Solar System$^2$ & $0.014\times 10^{-3}$ & \\
From emitting surface area$^3$  & &$0.9 \times 10^{-3}$ \\
From mass production rate$^4$  & & $0.3 \times 10^{-3}$\\
From integrating mass loss rate over 2 Gyr & & $5-10 \times 10^{-3}$ \\
\end{tabular}
\tablecomments{$^1$Stern (1996) for outer solar system and from 70 $\mu$m limit for HD69830; $^2$From Bidstrup et al. (2004) and Krasinsky et al. (2002); $^3$From scaling the emitting surface area to total mass ($\S 4.2$); $^4$Based on asteroid collisions for PR-drag dominated dust, $n_{gr} \propto N_c^2$ ($\S$4.4.2).} 
\end{table}

\end{document}